\documentclass[11pt]{article}
\pdfoutput=1
\usepackage{a4wide}
\usepackage{dcolumn}
\usepackage{bm}

\usepackage{graphicx}
\usepackage{amssymb,amsmath}
\usepackage{multirow}
\usepackage{cite}
\usepackage{color,url}
\usepackage[colorlinks=true
,urlcolor=blue
,anchorcolor=blue
,citecolor=blue
,filecolor=blue
,linkcolor=blue
,menucolor=blue
,linktocpage=true
,pdfproducer=medialab
,pdfa=true
]{hyperref}
\usepackage{amsmath, amsfonts, amssymb}
\usepackage{float}
\usepackage{pgf}
\usepackage[margin=2cm]{geometry}

\usepackage{epsfig,psfrag,rotating,soul}
\usepackage{rotfloat}
\usepackage{epsf}
\usepackage{graphics}
\usepackage{color}

\allowdisplaybreaks

\psfrag{lk}{$l_k$}
\psfrag{lm}{$l_m$}
\psfrag{blk}{$\bar {l}_k$}
\psfrag{blm}{$\bar {l}_m$}

\def\thefootnote{\fnsymbol{footnote}}

\newcommand{\gsim}{\stackrel{\scriptstyle >}{{ }_{\sim}}}

\newcommand{\TeV}{\ensuremath{\,{\rm TeV}}}
\newcommand{\GeV}{\ensuremath{\,{\rm GeV}}}

\newcommand{\Rk}{\ensuremath{R_{K}}}
\newcommand{\Rks}{\ensuremath{R_{K^{*0}}}}

\newcommand{\CnmuNP}{\ensuremath{C_9^{\rm{NP}\, \mu}}}
\newcommand{\CtmuNP}{\ensuremath{C_{10}^{\rm{NP}\, \mu}}}
\newcommand{\CntmuNP}{\ensuremath{C_{9,10}^{\rm{NP}\, \mu}}}

\newcommand{\CnpmuNP}{\ensuremath{C_9^{'\rm{NP}\, \mu}}}
\newcommand{\CtpmuNP}{\ensuremath{C_{10}^{'\rm{NP}\, \mu}}}

\newcommand{\CneNP}{\ensuremath{C_9^{\rm{NP}\, e}}}
\newcommand{\CteNP}{\ensuremath{C_{10}^{\rm{NP}\, e}}}
\newcommand{\CnpeNP}{\ensuremath{C_9^{'\rm{NP}\, e}}}
\newcommand{\CtpeNP}{\ensuremath{C_{10}^{'\rm{NP}\, e}}}

\newcommand{\CnlNP}{\ensuremath{C_9^{\rm{NP}\, \ell}}}
\newcommand{\CtlNP}{\ensuremath{C_{10}^{\rm{NP}\, \ell}}}

\newcommand{\Dms}{\ensuremath{\Delta M_s}}
\newcommand{\Dmsexp}{\ensuremath{\Delta M_s^{\rm{exp}}}}
\newcommand{\DmsSM}{\ensuremath{{\Delta M_s^{\rm{SM}}}}}
\newcommand{\Cll}{\ensuremath{C_{bs}^{LL}}}
\newcommand{\ACPmix}{\ensuremath{A_{CP}^{\rm{mix}}}}

\newcommand{\bsmumu}{\ensuremath{b\to s\mu^+\mu^-}}

\newcommand{\DchitSM}{\ensuremath{\Delta\chi^2_{\mathrm{SM}}}}
\newcommand{\sqDchitSM}{\ensuremath{\sqrt{\Delta\chi^2_{\mathrm{SM}}}}}
\newcommand{\Dchit}{\ensuremath{\Delta\chi^2}}

\newcommand{\chitdof}{\ensuremath{\chi^2_{\mathrm{min}}/\mathrm{d.o.f.}}}

\let\OLDthebibliography\thebibliography
\renewcommand\thebibliography[1]{
  \OLDthebibliography{#1}
  \setlength{\parskip}{0pt}
  \setlength{\itemsep}{0pt plus 0.3ex}
}

\begin{document}
\thispagestyle{empty}

\vspace*{2.5cm}

\vspace{0.5cm}

\begin{center}

\begin{Large}
\textbf{\textsc{Some results on Lepton 
Flavour Universality Violation}}
\end{Large}

\vspace{1cm}

{\sc
J. Alda$^{1}$%
\footnote{\tt \href{mailto:jalda@unizar.es}{jalda@unizar.es}}%
, J.~Guasch$^{2}$%
\footnote{\tt \href{mailto:jaume.guasch@ub.edu}{jaume.guasch@ub.edu}}%
, S.~Pe{\~n}aranda$^{1}$%
\footnote{\tt \href{mailto:siannah@unizar.es}{siannah@unizar.es}}%
}

\vspace*{.7cm}

{\sl
$^1$Departamento de F{\'\i}sica Te{\'o}rica, Facultad de Ciencias,\\
Universidad de Zaragoza, Pedro Cerbuna 12,  E-50009 Zaragoza, Spain

\vspace*{0.1cm}

$^2$Departament de F{\'\i}sica Qu{\`a}ntica i Astrof{\'\i}sica, Institut de Ci{\`e}ncies del Cosmos (ICCUB), \\
Universitat de Barcelona, Mart{\'\i}\ i Franqu{\`e}s 1, E-08028 Barcelona, Catalonia, Spain

}

\end{center}

\vspace*{0.1cm}

\begin{abstract}
\noindent
Motivated by recent experimental measurements on flavour physics, in tension
with Standard Model predictions, we perform an updated analysis of New Physics 
violating Lepton Flavour Universality, by using the effective Lagrangian
approach and in the $Z^{'}$ and $S_3$ leptoquark models.
We explicitly analyze the impact of considering complex Wilson coefficients 
in the analysis of $B$-anomalies, by performing 
a global fit of $\Rk$ and $\Rks$ observables, together with $\Dms$ and 
$\ACPmix$.  The inclusion of complex couplings
provides a slightly improved global
fit, and a marginally improved $\Dms$ prediction.

\end{abstract}

\def\thefootnote{\arabic{footnote}}
\setcounter{page}{0}
\setcounter{footnote}{0}

\newpage

\section{Introduction}
\label{intro}

At present, many interesting measurements on 
flavour physics are performed at the LHC~\cite{Aaij:2014pli,Aaij:2014ora,CMS:2014xfa,Aaij:2015esa,Aaij:2015yra,Aaij:2015oid,Aaij:2016flj,Aaij:2016cbx,Aaij:2017xqt,Aaij:2017vad,Aaij:2017vbb,Aaij:2017uff,TalkSimone,Aaij:2018jhg}.
Some relevant flavour transition processes
in order to constrain new physics at the LHC are the leptonic, semi-leptonic, baryonic and 
radiative exclusive decays. Some of these decays allow us to build optimized observables, 
as ratios of these decays, that are theoretically clean observables and whose measurements 
are in tension with Standard Model (SM) predictions. One example is the case of
observables in $b\rightarrow s l l$ transitions. Recently, the LHCb collaboration observed a
deviation from the SM predictions in the neutral-current $b \rightarrow s$
transition~\cite{Aaij:2014pli,Aaij:2014ora,Aaij:2015esa,Aaij:2015oid,Aaij:2016flj,Aaij:2017vbb,TalkSimone},
hinting at lepton flavour universality violation effects.
The results for ratios of branching ratios involving different lepton
flavours are given by~\cite{Aaij:2014ora,Aaij:2017vbb,TalkSimone},
\begin{eqnarray}
\label{eq:RK}
{\Rk=\frac{{\cal B}(B^+\to K^{+} \mu^+ \mu^-)}{{\cal  B}(B^+\to K^{+} e^+ e^-)}}&=&
0.745^{+0.090}_{-0.074}\pm 0.0036  \,\,\,\,1\GeV^{2}  < q^2 < 6\GeV^{2}\nonumber\\
{\Rks=\frac{{\cal B}(B^0\to K^{*0} \mu^+ \mu^-)}{{\cal  B}(B^0\to K^{*0} e^+ e^-)}}&=&
0.660^{+0.110}_{-0.070}\pm 0.024  \,\,\,\,\,\,\,0.045  \GeV^{2} < q^2 < 1.1  \GeV^{2}\nonumber\\
&& 0.685^{+0.113}_{-0.069} \pm 0.047  \,\,\,\,\,\,1.1 \GeV^{2} < q^2 < 6 \GeV^{2}
\end{eqnarray} 
where the first uncertainty is statistical and the second one comes from systematic effects.
In the SM $\Rk = \Rks = 1$ with theoretical uncertainties of 
the order of $1\%$~\cite{Hiller:2003js,Bordone:2016gaq}, as a consequence of Lepton Flavour Universality. 
The compatibility of the above results with respect to the SM
predictions is of $2.6 \,\sigma$ deviation in the first case and for $\Rks$,
in the low $q^{2}$ di-lepton invariant mass region is of about $2.3$ standard deviations; being in the
central$-q^{2}$ of $2.4 \,\sigma$. 
A discrepancy of about $3 \,\sigma$ is found when the
measurements of $\Rk$ and $\Rks$ are combined~\cite{Altmannshofer:2017fio}. 
Anomalous deviations were also observed in the angular distributions of the decay rate of
$B \to K^{*} \mu^+ \mu^-$, being the most significant discrepancy for the
$P_5^{'}$ observable~\cite{Aaij:2014pli,Aaij:2015oid}. 
The Belle Collaboration has also reported a discrepancy in angular observables consistent
with LHCb results~\cite{Wehle:2016yoi}. In addition, ATLAS and CMS collaborations 
have presented their updated results for the
angular parameters of the $B$ meson decay, 
$B^0\to K^{0} \mu^+ \mu^-$~\cite{ATLAS:2017dlmCONF,Carli:2017pby, ATLAS:2017dlm,Sirunyan:2017dhj}.

A great theoretical effort has been devoted to the understanding of these deviations, 
see for example~\cite{Hiller:2003js,Altmannshofer:2017fio,Altmannshofer:2013foa,Descotes-Genon:2014uoa,Hiller:2014yaa,Hiller:2014ula,Crivellin:2015lwa,Crivellin:2015era,Hurth:2016fbr,Capdevila:2017ert,Chobanova:2017ghn,Altmannshofer:2017wqy,Crivellin:2017zlb,Alok:2017jgr,Capdevila:2017bsm,Altmannshofer:2017yso,DAmico:2017mtc,Hiller:2017bzc,Geng:2017svp,Ciuchini:2017mik,Alok:2017jaf,Alok:2017sui} and references therein.
From the theoretical side, the ratios $\Rk$ and $\Rks$
are very clean observables; essentially free of hadronic uncertainties that cancel in the ratios~\cite{Hiller:2003js}. 
The experimental data has been used to constrain New Physics (NP) models. 
One useful way to analyze the effects of NP in these observables and
to quantify the possible deviations from the SM predictions is through the
effective Hamiltonian approach, allowing us a model-independent analysis
of new physics effects. In addition, one can compute this effective Hamiltonian in the context of
specific NP models. It has been shown that $Z^{'}$ and leptoquark models could explain the 
$\Rk, \Rks$ deviations. 

On the other hand, NP models are also severely constrained by other
flavour observables, for example in $B_{s}-$mixing. 
Recently an updated computation for the $B_{s}$ mesons mass difference in the SM 
has been presented~\cite{Bazavov:2016nty,Jubb:2016mvq,Buras:2016dxz,Kirk:2017juj,DiLuzio:2017fdq},
which shows a deviation with the experimental result~\cite{DiLuzio:2017fdq,Amhis:2016xyh}:
\begin{equation}
\label{eq:MSexp}
\Dmsexp= (17.757 \pm 0.021) \, {\rm ps}^{-1} \,,\,\,\,\, \DmsSM=(20.01
\pm 1.25) \, {\rm ps}^{-1}\,,
\end{equation}
such that $\DmsSM> \Dmsexp$ at about $2 \,\sigma$. 
This fact imposes additional constraints over the NP parameter space. 
Therefore, a global fit is mandatory when
considering all updated flavour observables. 
A negative contribution to  $\Dms$ is needed to reconcile it with the 
experimental result, in the context of some NP models (like $Z^{'}$ or leptoquarks)
it implies complex Wilson coefficients in the effective Hamiltonian of 
$\Rk, \Rks$~\cite{DiLuzio:2017fdq} (see also below).
To the best of our knowledge, most previous works have used only real Wilson coefficients 
in global fits of $\Rk$ and $\Rks$ observables 
  together with $\Delta M_s$, an exception being Ref.\cite{Alok:2017jgr}. An effect of introducing
    complex couplings is the generation of $CP$ asymmetries. The mixing-induced $CP$
    asymmetry in the $B$-sector can be measured through
$   \ACPmix\equiv\ACPmix(B_s\to J/\psi\phi)\equiv\sin(\phi_s^{c\bar{c}s})$, experimentally it is
measured to be~\cite{Amhis:2016xyh}:
\begin{equation}
\label{eq:ACPMixExp}
\ACPmix{}^{\rm \ exp}(B_s\to J/\psi\phi)=-0.021\pm0.031\ \ .
\end{equation}
In the SM it is given by 
 $\ACPmix{}^{\rm \ SM}=\sin(-2\beta_s)$\cite{DiLuzio:2017fdq,Artuso:2015swg,Lenz:2006hd}, 
with $\beta_s= 0.01852\pm 0.00032$\cite{Charles:2004jd} we obtain
$\ACPmix{}^{\rm \ SM} =-0.03703\pm0.00064$, which is consistent with
the experimental result~\eqref{eq:ACPMixExp} at the $\sim 0.5\,\sigma$ level.

Ref.\cite{Alok:2017jgr} performed
  fits for the $B$-decay physics observables using complex Wilson
  coefficients, in the model independent and model dependent
  approaches. The analysis of Ref.\cite{Alok:2017jgr} performs fits
  for the $B$-decay observables using complex couplings, without
  including the $\Dms$ or $\ACPmix$ observables, then
  Ref.\cite{Alok:2017jgr} proceeds to provide predictions to
  $CP$-violation observables. Ref.\cite{Alok:2017jgr} only includes
  $\Dms$ and $\ACPmix$ in the $Z'$-model fit. Our results agree with
  the ones of Ref.\cite{Alok:2017jgr} wherever comparable. 

The aim of the present work is to investigate the effects of complex
Wilson coefficients in the global analyses of NP in $B$-meson anomalies.
We assume a model independent 
effective Hamiltonian approach and we study the region of NP parameter space 
compatible with the experimental data, by considering the dependence of the 
results on the assumptions of imaginary and/or complex Wilson
coefficients. We compare our results with the case of considering only real Wilson
coefficients. A brief summary of the NP contributions to the effective Lagrangian
relevant for $b \to s \ell\ell$ transitions and $B_s$-mixing is presented in 
Section~\ref{sec:EFF}, where we also recall the need to consider complex Wilson 
coefficients in the analysis. In Section~\ref{sec:IMWCandRK} we discuss the effects
of having imaginary or complex Wilson coefficients on $\Rk$ observables. The impact of these
complex Wilson coefficients in the analysis of $B$-meson anomalies in two specific models,
$Z^{'}$ and leptoquarks, is included in Section~\ref{sec:Zandlepto}. We consider
a global fit of $\Rk$ and $\Rks$ observables, together with $\Delta M_s$ and 
$CP$-violation observable $\ACPmix$ in this analysis. Finally, 
conclusions are given in Section~\ref{conclusions}.

\section{Effective Hamiltonians and new physics models}
\label{sec:EFF}

The effective Lagrangian for $b \to s \ell\ell$ transitions is conventionally
given by~\cite{Buchalla:1995vs}, 
\begin{equation}
\label{eq:LagEff}
\mathcal{L}_\text{eff} = - \frac{4\,G_F}{\sqrt{2}} V_{tb}V_{ts}^*
\sum_{i, \ell}
(C_i^\ell O_i^\ell + C_i^{\prime\,\ell} O_i^{\prime\,\ell}) + \text{h.c.} \,,
\end{equation}
being $O_i^{(\prime)\,\ell}$ $(\ell = e,\mu)$ the operators and $C_i^{(\prime)\,\ell}$ the corresponding Wilson coefficients. 
The relevant semi-leptonic operators for explaining deviation in $\Rk$ observables, eq.~(\ref{eq:RK}), 
can be defined as,
\begin{align}
O_9^\ell &=\frac{e^2}{16\pi^2}
(\bar{s} \gamma_{\mu} P_{L} b)(\bar{\ell} \gamma^\mu \ell)\,,
& \!\!\!\!\!
O_9^{\prime \, \ell} &=\frac{e^2}{16\pi^2}
(\bar{s} \gamma_{\mu} P_{R} b)(\bar{\ell} \gamma^\mu \ell)\,,
\\
O_{10}^\ell &=\frac{e^2}{16\pi^2}
(\bar{s} \gamma_{\mu} P_{L} b)( \bar{\ell} \gamma^\mu \gamma_5 \ell)\,,
& \!\!\!\!\!
O_{10}^{\prime \, \ell} &=\frac{e^2}{16\pi^2}
(\bar{s} \gamma_{\mu} P_{R} b)( \bar{\ell} \gamma^\mu \gamma_5 \ell)\,.
\end{align}
The Wilson coefficients have contributions from the SM and
  NP,
$$
C_i^{(\prime)\,\ell} = C_i^{(\prime)\,\mathrm{SM}\, \ell} + C_i^{(\prime)\,\mathrm{NP}\, \ell}\ \ .
$$
In the present work we analyze the NP contributions
$C_i^{(\prime)\,\mathrm{NP}\, \ell}$.
In most of our analysis we will consider the left-handed
  Wilson coefficients $C_i^{\mathrm{NP}\, \ell}$, the
  righ-handed Wilson coefficients  $C_i^{\prime\,\mathrm{NP}\, \ell}$
  are treated briefly in the model-independent approach of
  Section~\ref{sec:IMWCandRK} (see Table~\ref{tableWC} below).
The NP contributions to $B_s$-mixing are described by the effective Lagrangian~\cite{Buchalla:1995vs}: 
\begin{equation}
\label{eq:Bslang}
\mathcal{L}^{\rm{NP}}_{\Delta B = 2} = - \frac{4 G_F}{\sqrt{2}} \left( V_{tb} V^*_{ts} \right)^2 
\left[ \Cll \left( \bar s_L \gamma_\mu b_L \right)^2 + \text{h.c.} \right] \,,
\end{equation}
where $\Cll$ is a Wilson coefficient. In order to study the allowed NP parameter space we follow
the same procedure as given in~\cite{DiLuzio:2017fdq}, comparing the experimental measurement of the mass difference with the prediction in the SM and NP. Therefore, the effects can be parametrized as~\cite{DiLuzio:2017fdq},
\begin{equation}
\label{eq:WCNP}
\frac{\Delta M_s}{\Delta M_s^{\rm{SM}}}
= \left| 1 + \frac{\Cll}{R^{\rm{loop}}_{\rm{SM}}} \right| \,, 
\end{equation}
where $R^{\rm loop}_{\rm SM} = 1.3397 \times 10^{-3}$~\cite{DiLuzio:2017fdq}.
The NP prediction to the $CP$-asymmetry $\ACPmix$ is
  given by\cite{DiLuzio:2017fdq,Artuso:2015swg,Lenz:2006hd}
\begin{equation}
  \label{eq:ACPMixNP}
\ACPmix=\sin(\phi_\Delta -2 \beta_s) \ \ \ , \ \ \
  \phi_\Delta={\rm Arg}\left(
    1+\frac{\Cll}{R^{\rm{loop}}_{\rm{SM}}}\right) \ \ \ ,
\end{equation}
where $\beta_s$ and $R^{\rm loop}_{\rm SM} $ have been given
  above.

 Since $\Dmsexp<\DmsSM$~(\ref{eq:MSexp}), eq.~\eqref{eq:WCNP} tells us
 that to obtain a prediction of $\Dms$ closer to $\Dmsexp$ the NP Wilson coefficient
 $\Cll$~\eqref{eq:Bslang} must be negative ($\Cll<0$). In a generic effective Hamiltonian approach, each Wilson coefficient is independent, and setting $\Cll<0$ has no effect on $\CnmuNP$, $\CtmuNP$, etc. However, explicit NP models give predictions on the Wilson coefficients which introduce correlations among them. We will concentrate on two specific models that have been proposed to solve the semi-leptonic $B_s$-decay anomalies: $Z'$ and leptoquarks.

We start with the $Z^{'}$ model that contains a $Z^{'}$ boson with mass $M_{Z^{'}}$ and whose extra NP operators 
can involve different chiralities. The part of the effective Lagrangian relevant for 
$\bsmumu$ transitions and $B_s$-mixing is given by~\cite{DiLuzio:2017fdq},
\begin{eqnarray}
\mathcal{L}_{Z'}^{\rm eff} &=& -\frac{1}{2 M^2_{Z'}} \left( \lambda^Q_{ij} \, \bar d_L^i \gamma_\mu d_L^j 
+ \lambda^L_{\alpha\beta} \, \bar \ell_L^\alpha \gamma_\mu \ell_L^\beta \right)^2 \\
&\sim& -\frac{1}{2 M^2_{Z'}}  \left[ 
(\lambda^Q_{23})^2 \left( \bar s_L \gamma_\mu b_L \right)^2 
+ 2 \lambda^Q_{23} \lambda^L_{22} (\bar s_L \gamma_\mu b_L) (\bar \mu_L \gamma^\mu \mu_L)
+ \text{h.c.} \right] +\cdots\,, \nonumber
\end{eqnarray}
where $d^i$ and $\ell^\alpha$ denote down-quark and charged-lepton mass eigenstates, and 
$\lambda^{Q}$ and $\lambda^{L}$ are hermitian matrices in flavour space. When matching the above equation 
with eqs.~(\ref{eq:LagEff}) and~(\ref{eq:Bslang}) one obtains the
expressions for the Wilson coefficients at the tree level~\cite{DiLuzio:2017fdq},
\begin{equation}
\label{C9C10Zp}
\CnmuNP = -\CtmuNP = - \frac{\pi}{\sqrt{2} G_F M^2_{Z'} \alpha} \left( \frac{\lambda^Q_{23} \lambda^L_{22}}{V_{tb} V^*_{ts}} \right) \, ,
\end{equation}
and 
\begin{equation} 
\label{CbsZp}
\Cll = \frac{\eta^{LL}(M_{Z'})}{4 \sqrt{2} G_F M^2_{Z'}} \left( \frac{\lambda^Q_{23}}{V_{tb} V^*_{ts}} \right)^2
\, , 
\end{equation}
where $\eta^{LL}(M_{Z'})>0$ encodes the running down to the bottom mass scale.   

From~(\ref{CbsZp}) it is clear that to obtain a negative $\Cll$ one
needs an imaginary number inside the square ($\lambda^Q_{23}/(V_{tb}
V^*_{ts})\in \mathbb{I}$), but this is the same factor that appears in 
$\CnmuNP=-\CtmuNP$ in~(\ref{C9C10Zp}). 
$\lambda^{L}_{22}\in\mathbb{R}$, since $\lambda$ is an hermitic matrix, 
then it follows that $\CntmuNP$ would be imaginary ($\CntmuNP\in\mathbb{I}$).
Of course, a purely imaginary coupling (or Wilson
  coefficient) is just a particular and extreme case of having a
  generic complex coupling. Once one abandons the restriction of
  considering real couplings it seems more natural to consider the
  most generic case of complex couplings. There is, however, a motivation
  to try also the extreme case of imaginary couplings: an
  imaginary $\lambda^Q_{23}/(V_{tb}V^*_{ts})$ provides a real
  $\Cll$~\eqref{CbsZp}, which in turn provides no
  additional contributions to the $CP$-asymmetry
  $\ACPmix$~\eqref{eq:ACPMixNP}, so imaginary couplings might
  provide a way of improving the predictions on $\Dms$ without
  introducing unwanted $CP$-asymmetries.

Now we focus on leptoquark models. Specifically, we consider the scalar leptoquark $S_3 \sim (\bar 3,3,1/3)$. The quantum number in
brackets indicate colour, weak and hypercharge representation,
respectively. The interaction Lagrangian reads~\cite{DiLuzio:2017fdq}
\begin{eqnarray}
\mathcal{L}_{S_3'} &=& -M_{S_3}^2 |S_3^a|^2 + y_{i\alpha}^{QL}
                       \overline{Q^c}^i (\varepsilon \sigma^a) L^\alpha
                       S_3^a+{\rm h.c.}\,\,,
\end{eqnarray}
where $\sigma^a$ are the Pauli matrices, $\varepsilon=i\sigma^2$, and
$Q^i$ and $L^\alpha$ are the left-handed quark and lepton doublets.
In this case, the contribution to the Wilson coefficients 
$\CntmuNP$ arises at the tree level and is 
given by~\cite{DiLuzio:2017fdq},
\begin{equation}
\label{C9C10LQ}
\CnmuNP = -\CtmuNP = 
\frac{\pi}{\sqrt{2} G_F M^2_{S_3} \alpha} \left( \frac{y_{32}^{QL} y_{22}^{QL*}}{V_{tb} V^*_{ts}} \right) \,.
\end{equation}
For $\Cll$ the contribution appears at the one loop level and can be
written as~\cite{Bobeth:2017ecx,DiLuzio:2017fdq}:
\begin{equation} 
\label{CbsLQ}
\Cll = \frac{\eta^{LL}(M_{S_3})}{4 \sqrt{2} G_F M^2_{S_3}} \frac{5}{64\pi^2} 
\left( \frac{\sum_{\alpha}y^{QL} _{3\alpha} y^{QL*} _{2\alpha}}{V_{tb} V^*_{ts}}
\right)^2 \,,
\end{equation}
where $\alpha=1,2,3$ is a lepton family index. Again, in order to
obtain $\Cll<0$ in~(\ref{CbsLQ}), the couplings must comply
$\sum_{\alpha=1}^3 y^{QL} _{3\alpha} y^{QL*} _{2\alpha}/({V_{tb} V^*_{ts}})\in \mathbb{I}$. 
If the combinations $ y^{QL} _{3\alpha} y^{QL*} _{2\alpha}/({V_{tb}
  V^*_{ts}})\in\mathbb{I}$, then the expression in eq.~\eqref{C9C10LQ}
suggests $\CntmuNP\in\mathbb{I}$. Of course, the
expression~\eqref{CbsLQ} is a sum over all generations, so it is
possible to set up a model with 
$y^{QL} _{32} y^{QL*} _{22}/({V_{tb}  V^*_{ts}})\in\mathbb{R}$, and to
have a cancellation such that the sum in eq.~\eqref{CbsLQ}
is imaginary, but this would be a highly fine-tuned scenario. If the
sum in~\eqref{CbsLQ} has an imaginary part, it would be most natural
if all its addends have some imaginary part. 

Here we have shown two examples of new physics models which justify the choice of imaginary (or complex) values for the Wilson coefficients $\CntmuNP$. In the next section we take an effective Hamiltonian approach and explore whether an imaginary or complex NP Wilson coefficients can accommodate the experimental $\Rk$ deviations. 

\section{Imaginary Wilson coefficients and $\Rk$ observables}
\label{sec:IMWCandRK}

Several groups have analyzed the 
predictions for the ratios~(\ref{eq:RK}) based on different global fits~\cite{Altmannshofer:2017fio,Hurth:2016fbr,Capdevila:2017ert, Alok:2017jgr,Capdevila:2017bsm,Altmannshofer:2017yso,DAmico:2017mtc,Hiller:2017bzc,Geng:2017svp,Ciuchini:2017mik},
extracting possible NP contributions or constraining it.
As it is well known, an excellent fit to the experimental data is obtained when 
$\CnlNP = - \CtlNP$; corresponding to left-handed lepton
currents. By considering this relation,
we investigate the effects of having imaginary Wilson coefficients on $\Rk$ observables.
For the numerical evaluation we use inputs values as given
in~\cite{Patrignani:2016xqp}.
The SM input parameters most relevant for our computation
  are:
\begin{eqnarray}
\label{eq:inputs}
&& \alpha_s(M_Z) = 0.1181(11)\,,\, 
G_F = 1.1663787(6)\times 10^{-5} \GeV^{-2}\,,\,\nonumber\\
&& M_W = 80.385(15)  \GeV\,, \,
{m}_t = 173.1(0.6)  \GeV\,,\nonumber\\
&& M_{B_s} = 5.36689(19)  \GeV\,,\nonumber\\
&& V_{tb} = 0.9991022\ \ , \ \  V_{ts} = -0.04137511-7.74823325\times 10^{-4}\;i \ \ \ ,
\end{eqnarray}
note that the product $V_{tb}V_{ts}^*$, which appears in the
  computation of Wilson coefficients in NP models~\eqref{C9C10Zp},
  \eqref{CbsZp}, \eqref{C9C10LQ}, \eqref{CbsLQ} is approximately a
  negative real number ($V_{tb}V_{ts}^*\simeq-0.04$).

\begin{figure}[t!]
\begin{center}
\begin{tabular}{cc}
\resizebox{0.45\textwidth}{!}{\includegraphics{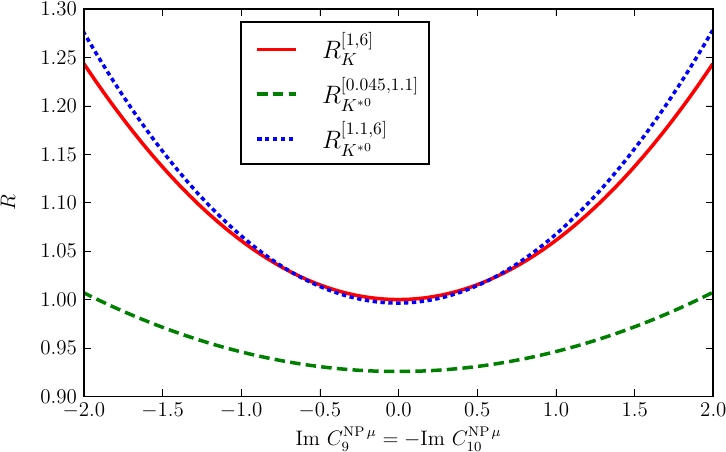}}&
\resizebox{0.45\textwidth}{!}{\includegraphics{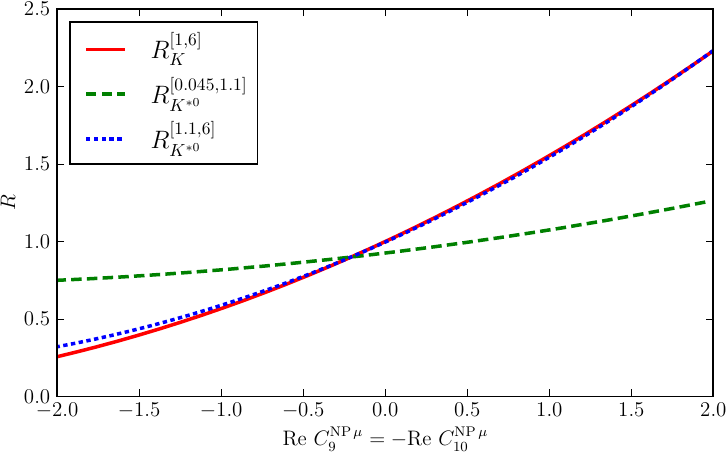}}\\
(a)&(b)
\end{tabular}
\caption{Values of $\Rk$ and $\Rks$ with (a) imaginary and (b) real Wilson
  coefficients.}
\label{im:RK_im}
\end{center}
\end{figure}

Figure~\ref{im:RK_im} shows the values of the ratios $\Rk$ and $\Rks$, in their respective 
$q^2$ ranges, when both Wilson coefficients $\CnmuNP$ and $\CtmuNP$ are
imaginary (Figure~\ref{im:RK_im}a) and when they are real (Figure~\ref{im:RK_im}b), by assuming that 
$\CnmuNP = -\CtmuNP$. If these two coefficients are imaginary, 
in all cases the minimum value for the ratio is obtained at the corresponding
SM point $\CnmuNP = -\CtmuNP =0$. The addition of non-zero imaginary Wilson coefficients results 
in larger values of $\Rk$ and $\Rks$, at odds with the experimental values 
$R_{K^{(*0)}}^{\mathrm{exp}} < R_{K^{(*0)}}^{\mathrm{SM}}$. 
This behaviour was already pointed out in
  Ref.~\cite{Hiller:2014ula}, where it is shown that the interference
  of purely imaginary Wilson with the SM vanishes, and therefore they
  can not provide negative contributions to $\Rk$, $\Rks$ (see also below).
In contrast,
as shown in the right panel,
values of $R_{K^{(*0)}} \sim 0.7$ (as in the experimental measurements)
are possible when the Wilson coefficients are real. 

\begin{figure}[t!]
\begin{center}
\begin{tabular}{cc}
\resizebox{0.52\textwidth}{!}{\includegraphics{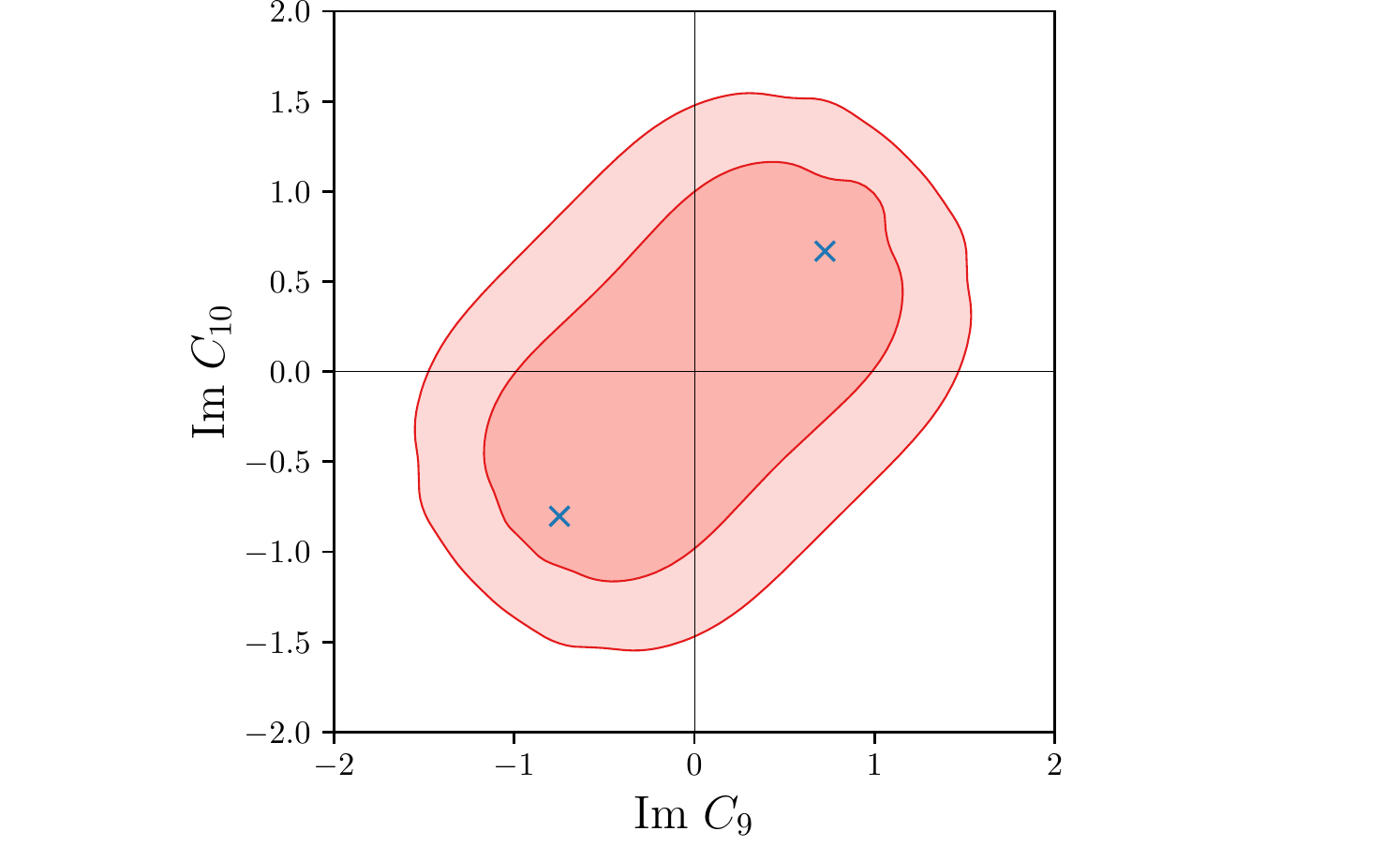}}&
\resizebox{0.52\textwidth}{!}{\includegraphics{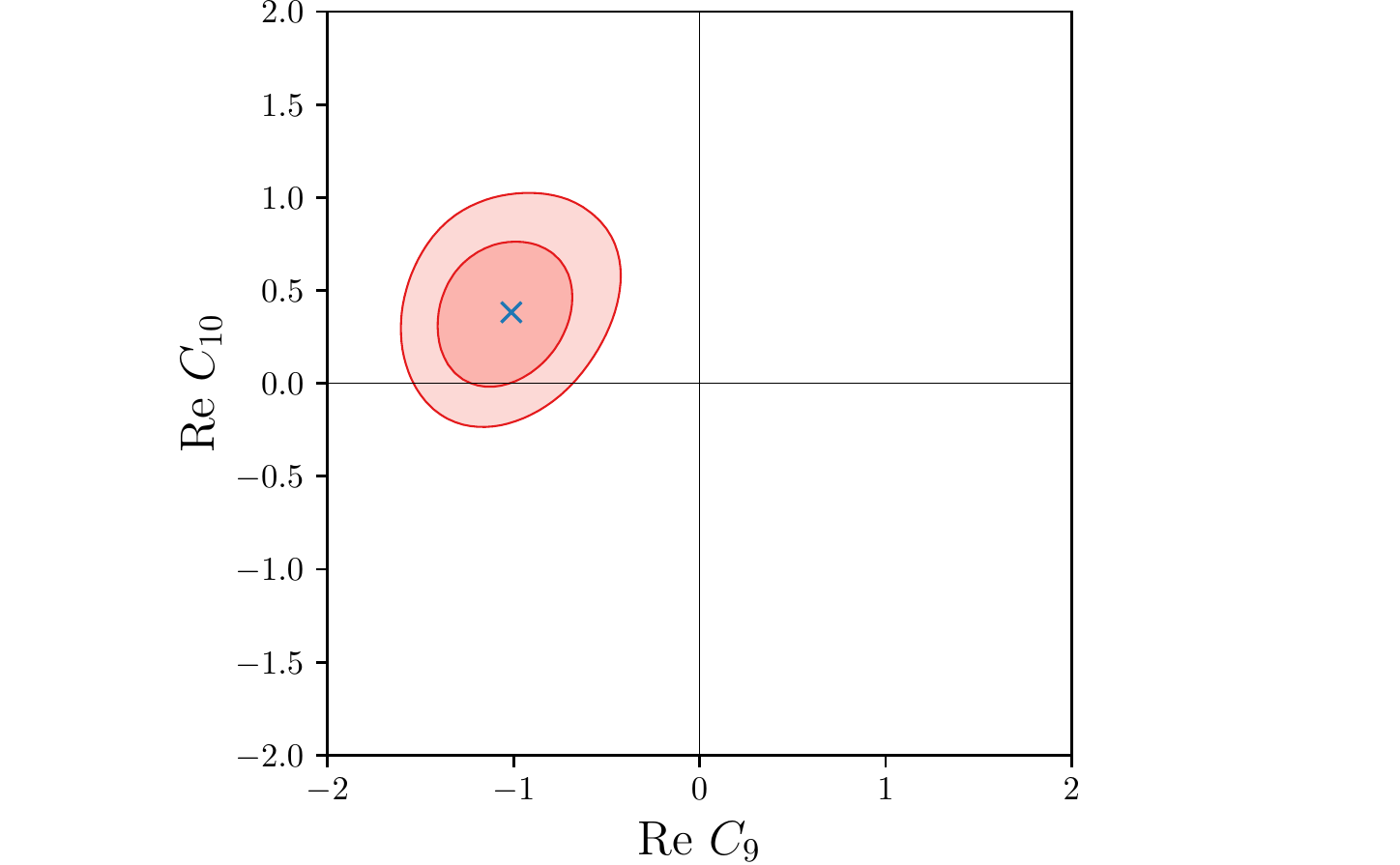}}\\
(a)&(b)
\end{tabular}
\caption{Best fit and $1 \,\sigma$ and $2 \,\sigma$ contours to semi-leptonic
  $B$-decays observables, $\Rk, \Rks, P_4'$ and $P_5'$, using (a)
  imaginary and (b) real Wilson coefficients.}
\label{im:fitimandre}
\end{center}
\end{figure}

We have done a global fit by including the ratios $\Rk$ and $\Rks$, 
and the angular observables $P_4'$ and
$P_5'$~\cite{Aaij:2015oid,ATLAS:2017dlmCONF,ATLAS:2017dlm,Sirunyan:2017dhj}\footnote{For the $P_4'$, $P_5'$
    observables we include all $q^2$ bins, except the ones
    around to the charm resonances $q^2\in [8.7,14]\GeV^2$,
    where the theoretical computation is not reliable. In total we
    include $15$ measurements for
    $P_4'$~\cite{Aaij:2015oid,ATLAS:2017dlmCONF,Sirunyan:2017dhj}
    and $21$ measurements for $P_5'$~\cite{Aaij:2015oid,ATLAS:2017dlmCONF,ATLAS:2017dlm,Sirunyan:2017dhj}.}.
Results are shown in Figure~\ref{im:fitimandre}. 
The allowed regions for imaginary values of $\CnmuNP$ and $\CtmuNP$ 
when fitting to measurements of a series of $\bsmumu$ observables are presented
in Figure~\ref{im:fitimandre}a, by assuming all other
Wilson coefficients to be SM-like. 
The numerical analysis has been done by using the open source code
{\it{flavio} 0.28}~\cite{flavio}, which computes the $\chi^2$ function with each $(\CnmuNP, \CtmuNP)$ pair.
The $\chi^2$ difference is evaluated with respect to the SM point, 
$\DchitSM= \chi^2_{\mathrm{SM}} - \chi^2_{\mathrm{min}}$. Then, the pull in $\sigma$
is defined as $\sqDchitSM$, in the case of only one Wilson coefficient, and
for the two-dimensional case it can be evaluated by using the inverse cumulative
distribution function of a $\chi^2$ distribution having
two degrees of freedom; for instance, $\Delta \chi^2 =2.29$ for $1\,\sigma$. 
The darker red shaded regions in Fig.~\ref{im:fitimandre}
  correspond to the points with
  $\Dchit=\chi^2-\chi^2_{\mathrm{min}}\leq 2.29$, that is, they
  are less than $1\,\sigma$ away from the best fit point, whereas the
  lighter red shaded regions correspond to $\Dchit\leq 6.18$ ($\equiv2\,\sigma$). 
The crosses mark the position of the best fit points.
In Fig.~\ref{im:fitimandre}a the $\chi^2$ function has a broad flat region centered around
  the origin, with two nearly symmetric minima found at ($\CnmuNP =
  0.72\,i$, $\CtmuNP = 0.74\;i$) and ($\CnmuNP=-0.75\,i$, $\CtmuNP=-0.74\,i$). The pull of the SM, 
defined as the probability that the SM scenario can describe the best fit assuming that 
$\DchitSM$ follows a $\chi^2$ distribution 
with 2 degrees of freedom, is of just $\sqDchitSM=1.42$
($\equiv 0.91\,\sigma$) and $\sqDchitSM=1.38$
($\equiv0.87\,\sigma$) respectively,  and both of them have the same $\chitdof=2.25$, that is,
purely complex couplings do not provide a good description of the data.  
For completeness, the fit to real values of the Wilson coefficients are 
included in Figure~\ref{im:fitimandre}b. 
Now the confidence regions are much tighter and do not include the SM point. In fact, 
the best fit point ($\CnmuNP = -1.09$, $\CtmuNP = 0.481$) improves the SM by
$\sqDchitSM=6.28$ ($\equiv 5.95\,\sigma$), and a much lower $\chitdof=1.24$.

Ref.~\cite{Hiller:2014ula} showed that imaginary Wilson
  coefficients do not interfere with the SM amplitude, an therefore
  imaginary $\CntmuNP$ can not decrease the prediction for $\Rk$,
  $\Rks$. This is numerically shown in the above analysis, where imaginary Wilson
  coefficients $\CntmuNP$ are not able to reduce significantly the prediction for
$R_{K}$, $\Rks$. To further investigate this question we have
analytically computed a numerical approximation to $\Rks$ as a
function of $\CnmuNP$, $\CtmuNP$ in the region $1.1 \leq q^2 \leq 6.0\GeV^2$. After integration and some approximations regarding
the scalar products of final state momenta, we obtain:
\begin{eqnarray}
\Rks \simeq \frac{0.9875+0.1759\, \mathrm{Re}\,\CnmuNP - 0.2954\,  \mathrm{Re}\, \CtmuNP + 0.0212|\CnmuNP|^2 + 0.0350 |\CtmuNP|^2}{1\,\  \ \ \ \ +0.1760\, \mathrm{Re}\,\CneNP - 0.3013\, \mathrm{Re}\,  \CteNP + 0.0212|\CneNP|^2 + 0.0357 |\CteNP|^2}
\ \ \ \ \ \ \ \ 
\nonumber\\ (1.1 \leq q^2 \leq 6.0\GeV^2). \label{RKform}
\end{eqnarray}
We have checked that this approximation reproduces the
  \textit{flavio}-computed value of $\Rks$ to better than $4\%$ in
a large region of the parameter space. Now, if we assume that NP does
not affect the electron channel ($\CneNP=\CteNP=0$), it is clear that
to obtain $\Rks<\Rks^{\mathrm{SM}}$ one needs to introduce $\CnmuNP$
and $\CtmuNP$ with a non-zero real part: the only possible
negative contributions come from the $\mathrm{Re}\,\CnmuNP$,
$\mathrm{Re}\,\CtmuNP$ terms, whereas the $|\CnmuNP|^2$,
$|\CtmuNP|^2$ terms have a positive-defined sign, and can not
reduce the value of $\Rks$. Thus, purely imaginary values of
$\CntmuNP$ contribute only to the modulus (positive-definite) and
not to the real part, and can not bring the prediction of $\Rks$
closer to the experimental value. In addition, this expression tells
us that the better option to reduce the prediction of $\Rks$ is
using a real negative $\CnmuNP$, and a real positive
$\CtmuNP$. This is actually the result that we have obtained in
our numerical analysis. Fig.~\ref{im:RK_im}b shows that, for real
Wilson coefficients, the lowest prediction for \Rks\  is obtained for $\CnmuNP=-\CtmuNP<0$, and
Fig.\ref{im:fitimandre}b shows that the best fit is obtained for
negative $\CnmuNP$ and positive $\CtmuNP$. Fig.~\ref{im:RK_im}a
shows that, in general, imaginary Wilson coefficients give positive
contributions to \Rk, \Rks, in accordance with eq.~\eqref{RKform}. Of
course, the full expression is richer than eq.~\eqref{RKform}, and
we expect some deviations, Fig.~\ref{im:fitimandre}a shows that the
best fit point is not the SM ($\CnmuNP=\CtmuNP=0$), but the best
fit regions are centered around it,
and the SM pull with respect the best fit points is small.
\begin{table}[t]
\centering
\begin{tabular}{|c|c|c|c|c|}\hline
& Best fit(s) & Pull ($\sqDchitSM$) & Pull ($\sigma$)   & $\chitdof$ \\\hline
$\CnmuNP$ & $-1.11-0.02\;i$ & 5.94 & 5.60 $\sigma$ & 1.35 \\\hline
\multirow{2}{*}{$\CtmuNP$} & $1.66+1.99\;i$ & \multirow{2}{*}{5.02} &  \multirow{2}{*}{4.65 $\sigma$} &  \multirow{2}{*}{1.62}  \\ & $1.67-2.01\;i$ & & & \\\hline
\multirow{2}{*}{$\CnmuNP = -\CtmuNP$} & $-1.16+1.14\;i$ & \multirow{2}{*}{6.06} & \multirow{2}{*}{5.72 $\sigma$} & \multirow{2}{*}{1.31} \\ & $-1.18-1.18\;i$ & & & \\\hline
$\CnpmuNP$  & $-0.24-0.003\;i$ & 1.07 & 0.57 $\sigma$& 2.27 \\\hline
$\CtpmuNP$ & $0.33-0.014\;i$ & 2.22 & 1.72 $\sigma$& 2.17  \\\hline 
\multirow{2}{*}{$\CneNP$} & $-3.29+5.02\;i$ & \multirow{2}{*}{4.85} & \multirow{2}{*}{4.47 $\sigma$} & \multirow{2}{*}{1.67} \\ & $-3.35-5.04\;i$ & & & \\\hline
\multirow{2}{*}{$\CteNP$} & $-0.27+3.48\;i$ & \multirow{2}{*}{4.72} & \multirow{2}{*}{4.34 $\sigma$} & \multirow{2}{*}{1.70} \\ &  $-0.27-3.48\;i$ & & & \\\hline
\multirow{2}{*}{$\CneNP = -\CteNP$} & $-3.29+4.58\;i$ & \multirow{2}{*}{4.85} & \multirow{2}{*}{4.47 $\sigma$} & \multirow{2}{*}{1.67} \\ & $-3.35-4.59\;i$ & & & \\\hline
\multirow{2}{*}{$\CnpeNP$} & $-0.59+3.89\;i$ & \multirow{2}{*}{4.81} & \multirow{2}{*}{4.43 $\sigma$} & \multirow{2}{*}{1.68} \\ & $-0.59-3.89\;i$ & & & \\\hline
\multirow{2}{*}{$\CtpeNP$} & $0.52+3.88\;i$ & \multirow{2}{*}{4.81} & \multirow{2}{*}{4.43 $\sigma$} & \multirow{2}{*}{1.68} \\ & $0.53-3.88\;i$ & & &  \\\hline
\end{tabular}
\caption{Best fit Wilson coefficients complex values to semi-leptonic decay observables
$\Rk, \Rks, P_4'$ and $P_5'$, allowing only one free coefficient at a
time. Shown are also the corresponding pulls, and  $\chitdof$.}
\label{tableWC}
\end{table}

\begin{table}
\centering
\begin{tabular}{|c|c|c|c|}\hline
 & \Rk & $\Rks^{[0.045,1.1]}$ & $\Rks^{[1.1,6]}$ \\\hline
$\CnmuNP$  & $ 0.77 \pm 0.03 $ & $ 0.887 \pm 0.009 $ & $ 0.82\pm 0.04 $ \\\hline
$\CtmuNP$ & $ 0.78 \pm 0.05 $ & $ 0.87 \pm 0.03 $ & $ 0.80 \pm 0.10 $ \\\hline
$\CnmuNP = -\CtmuNP$ & $0.59 \pm 0.08$ & $0.83 \pm 0.03$ & $0.63 \pm 0.09$ \\\hline
$\CnpmuNP$  & $0.95 \pm 0.05$ & $ 0.96\pm 0.03$ & $1.09 \pm 0.09$ \\\hline
$\CtpmuNP$ & $ 0.92\pm 0.07$ & $0.95 \pm 0.03$ & $1.07 \pm 0.09$ \\\hline
$\CneNP$  & $0.76 \pm 0.09$ & $0.69 \pm 0.12$ & $0.52 \pm 0.17$ \\\hline
$\CteNP$  & $0.69 \pm 0.06$ & $0.77 \pm 0.06$ & $0.59 \pm 0.13$ \\\hline
$\CneNP = -\CteNP$ & $0.76 \pm 0.09$ & $0.70 \pm 0.10$ & $0.52 \pm 0.17$ \\\hline
$\CnpeNP$  & $0.75 \pm 0.09$ & $0.71 \pm 0.10$ & $0.52 \pm 0.18$ \\\hline
$\CtpeNP$ & $0.75 \pm 0.09$ & $0.80 \pm 0.09$ & $0.66 \pm 0.14$ \\\hline
\end{tabular}
\caption{$\Rk$, $\Rks$ predictions with $1\,\sigma$ uncertainties
  corresponding to the best fit Wilson coefficients of Table~\ref{tableWC}.}
\label{tableWC_predictions}
\end{table}

We conclude  that, actually, a NP explanation for \Rk, \Rks\
requires that \CnmuNP, \CtmuNP\  have a non-zero real part, whereas we
saw above that NP explanation for \Dms\  requires that \CnmuNP, \CtmuNP\
have a non-zero imaginary part. Then, to have a NP explanation for
both observables \CnmuNP, \CtmuNP\  should be general complex numbers.
Following this reasoning we have performed a global fit to the semi-leptonic decay observables
$\Rk, \Rks, P_4'$ and $P_5'$ using generic complex Wilson
coefficients allowing only one free Wilson coefficient at a time. 
Table~\ref{tableWC} shows the best fit values, pulls (defined
  as $\sqDchitSM$) and $\chitdof$, for scenarios with
NP in one individual complex Wilson coefficient, and
Table~\ref{tableWC_predictions} shows the prediction for $\Rk$, $\Rks$
for the corresponding central values of each fit, together with the $1\,\sigma$ uncertainties. The primed Wilson
coefficients are also included. We found that the best fit of $\Rk$ and $\Rks$ and 
the angular distributions is obtained for $\CnmuNP =
  -1.11-0.02 \;i$, for $\CtmuNP$ we find two points with similar
minimum value for $\chi^2$ with opposite signs of the imaginary part, $\CtmuNP=1.66+1.99\;i$  and
$\CtmuNP=1.65-2.10\;i$. Assuming $\CnmuNP=-\CtmuNP$ we also obtain a
double minimum $\CnmuNP=-\CtmuNP=-1.16+1.14\;i$ and
$\CnmuNP=-\CtmuNP=-1.18-1.18\;i$ with a pull of $\sqDchitSM=6.06$
($\equiv 5.72\,\sigma$) and a $\chitdof=1.31$.
By looking at $\chitdof$ we see that the scenarios with
only $\CnmuNP$ or $\CnmuNP=-\CtmuNP$ provide the best description of
experimental data, whereas the scenarios with $\CnpmuNP$ and
$\CtpmuNP$ provide the worst description.
If only real Wilson coefficients are chosen the best fit of $\Rk$ and $\Rks$ yields
$\CnmuNP = -1.59$, $\CtmuNP=1.23$ or $\CnmuNP = -\CtmuNP=-0.64$, with
a pull around $4.2 \,\sigma$~\cite{Altmannshofer:2017yso}.

Ref.\cite{Alok:2017jgr} also provides fits for complex
  generic Wilson coefficients. Their \textit{scenario I} corresponds to
  our first line in Table~\ref{tableWC}, our best fit value agrees
  with their result ($\CnmuNP= (-1.1\pm0.2) + (0\pm0.9\,i)$), within the large uncertainties they give for the
  imaginary part, but we obtain larger pulls ($5.6\,\sigma$
  vs. $4.2\,\sigma$ of Ref.\cite{Alok:2017jgr}). Their
  \textit{scenario II} corresponds to our third line in
  Table~\ref{tableWC} ($\CnmuNP=-\CtmuNP$), we agree with the main
  features of their fit, for the real part they obtain $\mathrm{Re}(\CnmuNP)=\mathrm{Re}(\CtmuNP)=-0.8\pm0.3$, we obtain a slightly
  smaller real part, but they agree within uncertainties, both of us
  obtain a double minimum for the imaginary part $\sim \pm
  (1.1-1.2)\,i$, again, we obtain a slightly larger pull
  ($5.72\,\sigma$ vs. $4.0$, $4.2\,\sigma$ of Ref.\cite{Alok:2017jgr}).  

Choosing complex Wilson coefficients also implies additional constraints from $CP$-violating
observables. This fact has not been considered in the previous analysis. In the next section we study
the consequences of having these coefficients in the analysis of
$B$-meson anomalies on some NP models and we consider 
a global fit of both the ratios $\Rk$ and $\Rks$ and the angular observables $P_4'$ and $P_5'$, and also 
the $CP$-mixing asymmetry.

\section{$B_s$-mixing and NP models}
\label{sec:Zandlepto}

Several NP models that are able to explain the lepton flavour universality violation effects are
constrained by other flavour observables like $B_s$-mixing. In particular the parameter space 
of $Z^{'}$ and leptoquark models are severely constrained by the present experimental results of 
$\Delta M_s$~\cite{DiLuzio:2017fdq}. Besides, as already mentioned, additional constraints
emerge from $CP$-violating observables when considering complex couplings.
Ref.\cite{DiLuzio:2017fdq} argues that nearly imaginary Wilson coefficients could explain 
the discrepancies with the $\Delta M_s$ experimental measurement, but a global fit of $\Rk$ and $\Rks$ 
observables, together with $\Delta M_s$ and $CP$-violation observable $\ACPmix$ in
$B_{s}\rightarrow J/\psi\phi$ decays should be performed. In the next subsections we investigate these 
issues for the case of $Z^{'}$ and leptoquark models. 

\subsection{ $Z^{'}$  fit}

From now on, a global fit of $\Rk$ and $\Rks$ 
observables, $\Delta M_s$ and the $CP$-violation observable
$\ACPmix$ is included in our analysis. 

\begin{figure}[t]
\centering
\resizebox{0.6\textwidth}{!}{\includegraphics{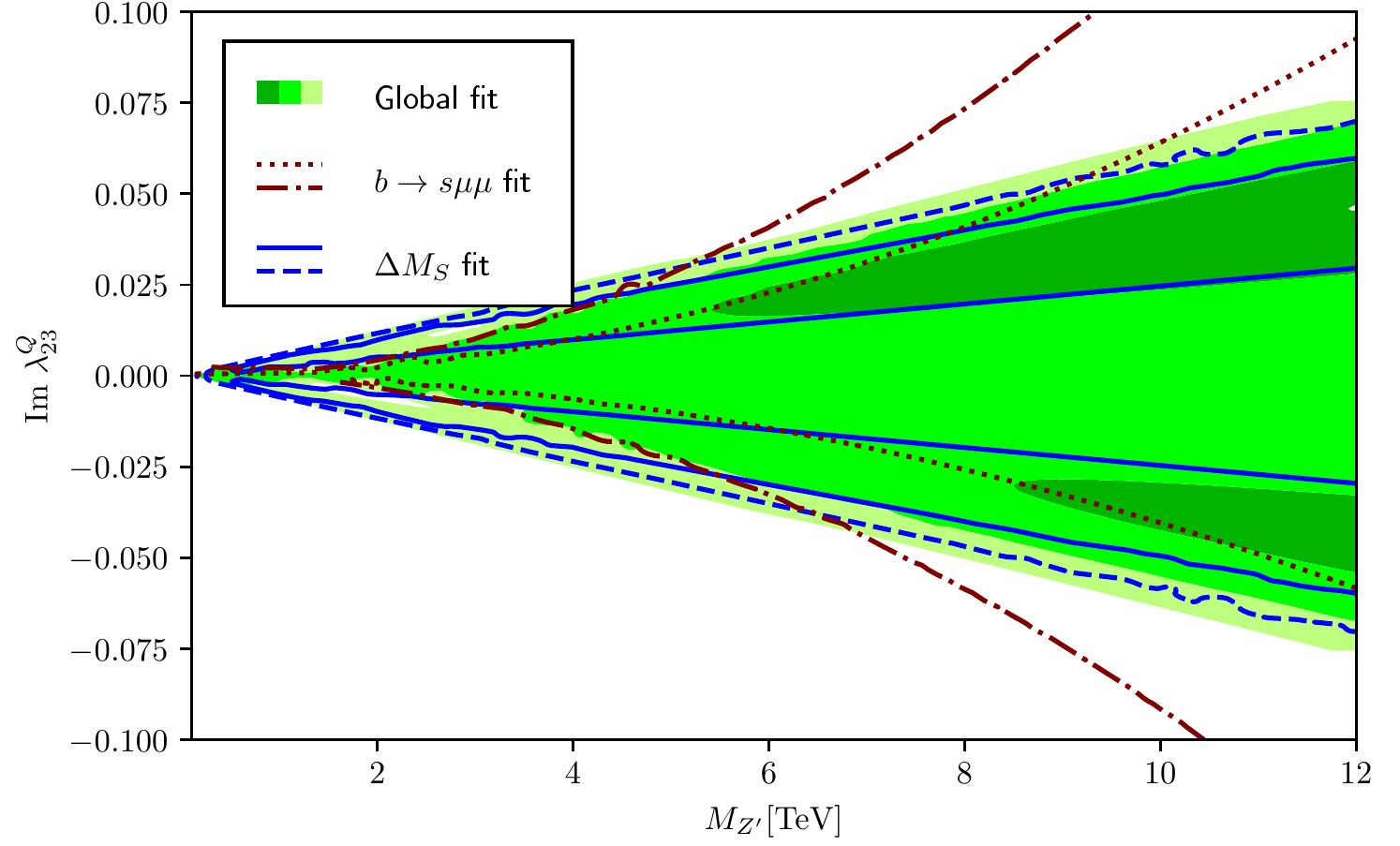}}
\caption{Fit on $Z'$ parameter space in the
  $M_{Z'}$-Im $\lambda_{23}^Q$ plane (see text).}
\label{im:WCZ}
\end{figure}

Figure \ref{im:WCZ} shows the fits on the $Z'$ mass $M_{Z'}$ and the
imaginary coupling $\lambda_{23}^Q$ (setting $\lambda_{22}^L=1$) 
imposed by $\bsmumu$ decays and $B_s$-mixing. The red lines
(dotted, dash-dotted) correspond to the fit using only semi-leptonic
$B$-meson decays, i.e. $\bsmumu$ as in Figure
\ref{im:fitimandre} plus the branching ratios $\mathrm{BR}(B_s\to
\mu^+ \mu^-)$ and $\mathrm{BR}(B^0 \to \mu^+ \mu^-)$. The best fit
region is the one between the curves; dotted lines: $\Dchit = 1$, dash-dotted lines: $\Dchit = 4$. Blue lines
(solid, dashed) correspond to the fit to $B_s$-mixing observables
$\Delta M_s$ and $\ACPmix$. The best fit region is the one between the
lines; solid lines $\Dchit = 1$, dashed lines $\Dchit =
4$,  there are two
  regions with $\Dchit<1$, but between them $\Dchit$ is
  always smaller than 4. The green regions are the combined global fit: dark region
$\Dchit \leq 1$, medium $\Dchit\leq4$ and light $\Dchit \leq 9$.

\begin{table}[t]
\centering
\begin{tabular}{|c|c|c|c|}\hline
Best fits 	&	Real 	&	Imaginary 	&	Complex 	\\\hline
 $\lambda_{23}^{Q}$ 	&	 $-0.002 $ 	&	 $\pm 0.047\;i$ 	&	 $-0.0020-0.0021\;i$ 	\\\hline
 $M_{Z'}$ 	&	1.31  TeV 	&	12  TeV 	&	1.08  TeV 	\\\hline
   Pull ($\sqDchitSM$) 	&	 5.70 	&	1.61	&	 6.05	\\\hline
  Pull ($\sigma$) 	&	5.39 $\sigma$ 	&	1.09 $\sigma$	&	5.43 $\sigma$ 	\\\hline
 $\chitdof$  & 	1.41		&	2.12 	& 1.34		\\\hline
 $\Rk$ 	&	$0.66 \pm 0.05$	&	$ 1.00 \pm 0.01 $	&	$0.65 \pm 0.07 $	\\\hline
 $\Rks^{[0.045,1.1]}$ 	&	$0.849 \pm 0.013$	&	$ 0.93 \pm 0.02$	&	$ 0.84\pm 0.02$	\\\hline
 $\Rks^{[1.1, 6]}$ 	&	$0.68 \pm 0.05$	&	$1.00 \pm 0.01$	&	$ 0.68\pm 0.07$	\\\hline
 $\Delta M_s$	&	 $ 20.41 \pm 1.26 $ ps${}^{-1}$ 	&	 $18.0 \pm 1.7 $ ps${}^{-1}$ 	&	  $ 19.95\pm 1.27$ ps${}^{-1}$	\\\hline
$\ACPmix$	&	$-0.0369 \pm 0.0002$	&	$ -0.041 \pm 0.002 $	&	$-0.035 \pm 0.003$	\\\hline
\end{tabular}
\caption{Best fits, and corresponding pulls, to $\Rk, \Rks, \Delta M_s$ 
and $\ACPmix$; considering real,
imaginary and complex Wilson coefficients on the $Z^{'}$
model. Shown are also the corresponding pulls,
$\chitdof$, and the predictions for
  semi-leptonic decay observables $\Rk$, $\Rks$;  $\Dms$ and $\ACPmix$
  with $1\,\sigma$ uncertainties.}
\label{tableZ}
\end{table}

\begin{figure}[t]
\centering
\resizebox{0.6\textwidth}{!}{\includegraphics{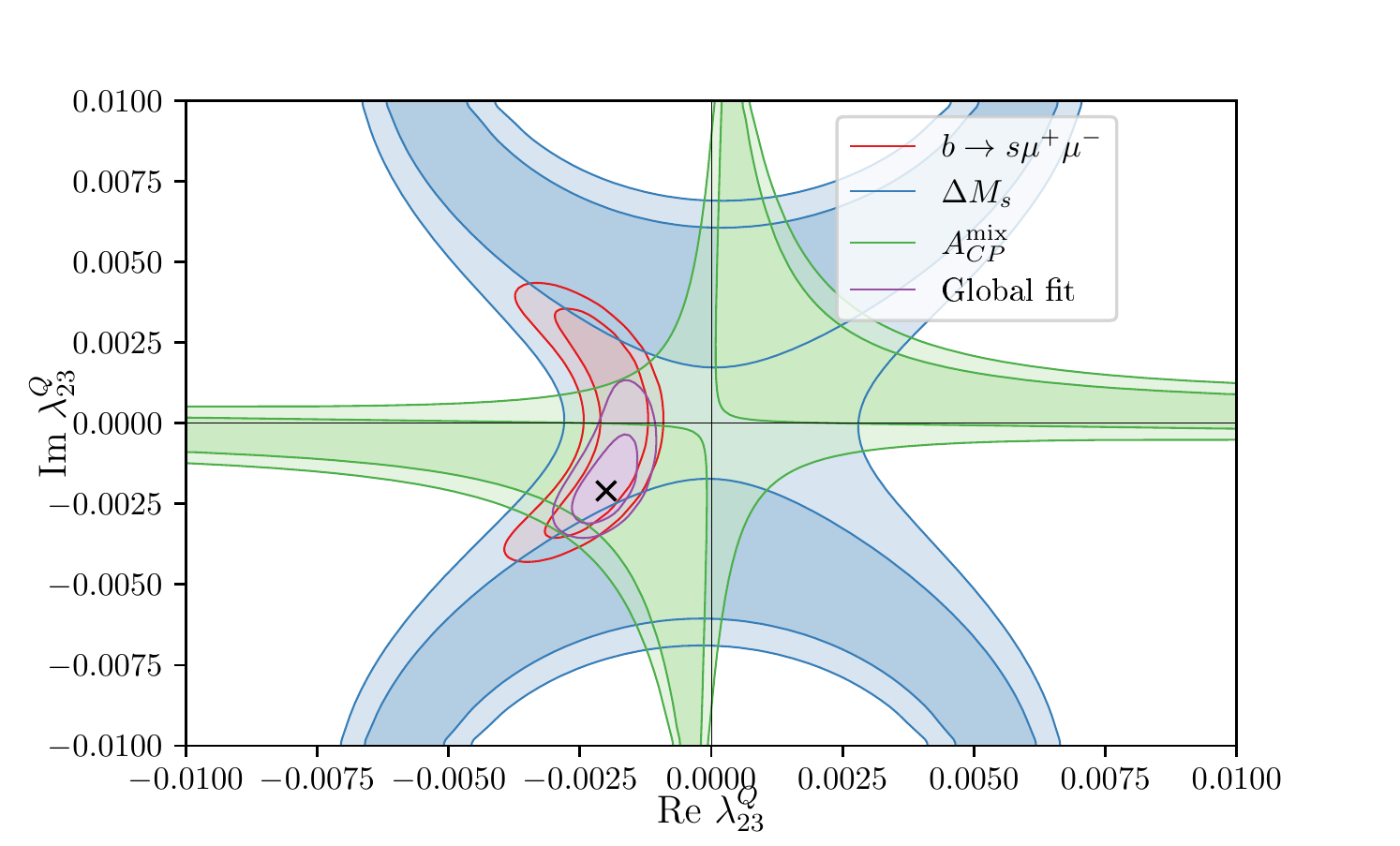}}
\caption{Fit on $Z'$ parameter space in the
  $\lambda_{23}^Q$ complex plane for the best fit $Z'$
  mass $M_{Z'}=1.08\TeV$ (see text).}
\label{fig:fitcompl_Z}
\end{figure}

The best fit for the $\bsmumu$ observables in the region
  under study is $M_{Z'}= 11\TeV$, $\lambda_{23}^Q = 0.015\;i$, with a
  tiny $\sqDchitSM=0.23$, which makes it statistically
  indistinguishable from the SM, and a large $\chitdof=2.92$
  which indicates that it does not provide a good fit to the data.
 For the $B_s$-mixing observables, the best fit is found at 
the maximum allowed mass $M_{Z'}= 12\TeV$, $\lambda_{23}^Q =\pm 0.05\;i$, 
which corresponds to $\Cll= -1.54\times 10^{-4}$. 
The SM has a pull of $\sqDchitSM = 1.73$ ($\equiv1.21\,\sigma$), and
the minimum has a $\chitdof=0.52$.
The best fit when all observables are considered, in the 
$M_{Z'}$ region of our analysis, and $\lambda_{23}^Q$ being a
pure imaginary coupling, is found at 
$M_{Z'} = 12\TeV$, $\lambda_{23}^Q = \pm0.047\;i$, 
and the pull of the SM is $\sqDchitSM = 1.61 (\equiv
1.09\,\sigma)$ and $\chitdof=2.12$. 
Larger values of $M_{Z'}$ do not improve the pull of the
  SM. Actually, if one allows larger values for $M_{Z'}$ the best fit
  point has a linear relation between the coupling and the maximal
  allowed mass: 
  $\lambda_{23}^Q\simeq i\,(3.95\times
  M_{Z'}^{\mathrm{max}}/\TeV)\times 10^{-3}$. 
 This linear relation produces a (approximately) constant
  $\Cll$~\eqref{CbsZp}, with a $\Dms$
  prediction close to the experimental value~\eqref{eq:MSexp}, while the contributions
  to $|\CntmuNP|$ decrease as $M_{Z'}^{-1}$~\eqref{C9C10Zp}. Since
  imaginary couplings worsen the $\Rk$, $\Rks$ prediction, the larger $M_{Z'}$
  provides better predictions for them, bringing them closer to the SM value. The best fit $\DchitSM$
  grows very slowly with growing allowed $M_{Z'}$. 
Table~\ref{tableZ} summarizes the best fit values for  $\lambda_{23}^{Q}$ and
$M_{Z'}$, and corresponding pulls, to $\Rk$ and $\Rks$ 
observables, $\Delta M_s$ and $\ACPmix$; considering real,
imaginary and complex Wilson coefficients. Results for the above observables 
in each scenario are included in this table. It is clear that $\Rk$ and $\Rks$ 
observables prefer real Wilson coefficients, as expected. For real
couplings the description is better than the SM, with a pull of
$5.39\,\sigma$ but it does not improve the prediction for
$\Dms$. Contrary, to improve the prediction for $\Dms$
imaginary couplings are required in the $Z^{'}$ model,
however the pull with respect the SM is small, and it has a
  large $\chitdof$. When allowing
generic complex couplings (third column in Table~\ref{tableZ}) we find
that the best fit point is close to the best fit point using only real
couplings (first column in Table~\ref{tableZ}), and the pull with
respect the SM improves slightly ($5.43\,\sigma$ versus $5.39\,\sigma$),
and the predictions for the observables are also close to the pure real
couplings case, showing a slight improvement in the
prediction for $\Dms$.

Fig.~\ref{fig:fitcompl_Z} shows the best fit regions in the
  complex $\lambda_{23}^Q$ plane for the best fit mass value
  $M_{Z'}=1.08\TeV$ (Table~\ref{tableZ}). The red region shows the
  2-dimensional 1
  and 2-$\sigma$ allowed values ($\Dchit=2.29$, $6.18$) including only the $\bsmumu$
  observables, the blue region shows the 1 and 2-$\sigma$ allowed
  values including only $\Dms$, and the green region show the 1 and
  2-$\sigma$ allowed values including only $\ACPmix$, the violet
  region shows the combined fit. Here we see the tension between the
  $\bsmumu$ and $\Dms$ fits. $\bsmumu$ selects a region
  around the real axis of the coupling, whereas $\Dms$ selects regions
  away from it. There are two small intersection regions for the
  1-$\sigma$ allowed values of both fits. The $\ACPmix$ fit selects
  one of these regions, and breaks the degeneracy. Actually, the
  $\bsmumu$ fit selects fixed values of $\CnmuNP=-\CtmuNP$,
  eq.~\eqref{C9C10Zp}, since $\CnmuNP=-\CtmuNP$ scale as $\sim
  \lambda_{23}^Q/M_{Z'}^2$, for fixed $\CnmuNP=-\CtmuNP$ the allowed
  values of $\lambda_{23}^Q$ (red region in Fig.~\ref{fig:fitcompl_Z})
  around the real axis will grow as $M_{Z'}^2$, but, at the same time,
  the allowed region will move away from the imaginary axis as
  $M_{Z'}^2$. On the other hand, the fit on $\Dms$ selects fixed values of
  $\Cll$, eq.~\eqref{CbsZp}, since $\Cll\sim
  (\lambda_{23}^Q)^2/M_{Z'}^2$, for fixed $\Cll$ the 1-$\sigma$
  unfavored region around the origin (light blue region in
  Fig.~\ref{fig:fitcompl_Z}) will grow as $\lambda_{23}^Q\sim
  M_{Z'}$. As $M_{Z'}$ grows, the red region moves away from the
  origin as $M_{Z'}^2$, but the blue region expands only as $M_{Z'}$,
  so that at some $M_{Z'}$ value their 1-$\sigma$ regions do not
  longer intersect. This is the reason why we obtain a relatively low
  $M_{Z'}$ in the fits of Table~\ref{tableZ}.

Ref.\cite{Alok:2017jgr} provides also a fit for the $Z'$
  model, using a fixed $M_{Z'}=1\TeV$, this value is close to our best
fit value of Table~\ref{tableZ}. For $\lambda_{22}^L=1$ they obtain
the best fit coupling $\lambda_{23}^Q= (-0.8\pm0.3)\times 10^{-3} +
(-0.4 \pm 3.1)\times 10^{-3}\,i$ with a pull of $4.0\,\sigma$. Our
best fit values agree with them within uncertainties. Note that we do
not provide uncertainties for the best fit values, the reason being
that the parameters are not independent, the 2-dimensional best
fit regions in Fig.~\ref{fig:fitcompl_Z} are not ellipses, and the
best fit points are not on the center of the figures, so that giving a
central value with 1-dimensional uncertainties overestimates the
uncertainty and leads to confusion about the meaning and position of
the best fit point.

We conclude that, in the framework of
$Z'$ models, $\Rk-\Rks$ observables are better described than in the
SM, with a pull $\gsim 5.39\,\sigma$ for $M_{Z'}\simeq 1-1.3 \TeV$, and
a coupling with a real part ${\rm Re}(\lambda_{23}^Q)\simeq -0.002$. The
presence of a similar imaginary part for the coupling
${\rm Im}(\lambda_{23}^Q)\simeq -0.0021$ improves slightly the fit,
as well as the $\Dms$ prediction.

\subsection{Leptoquark fit}

\begin{figure}[t]
\centering
\resizebox{0.6\textwidth}{!}{\includegraphics{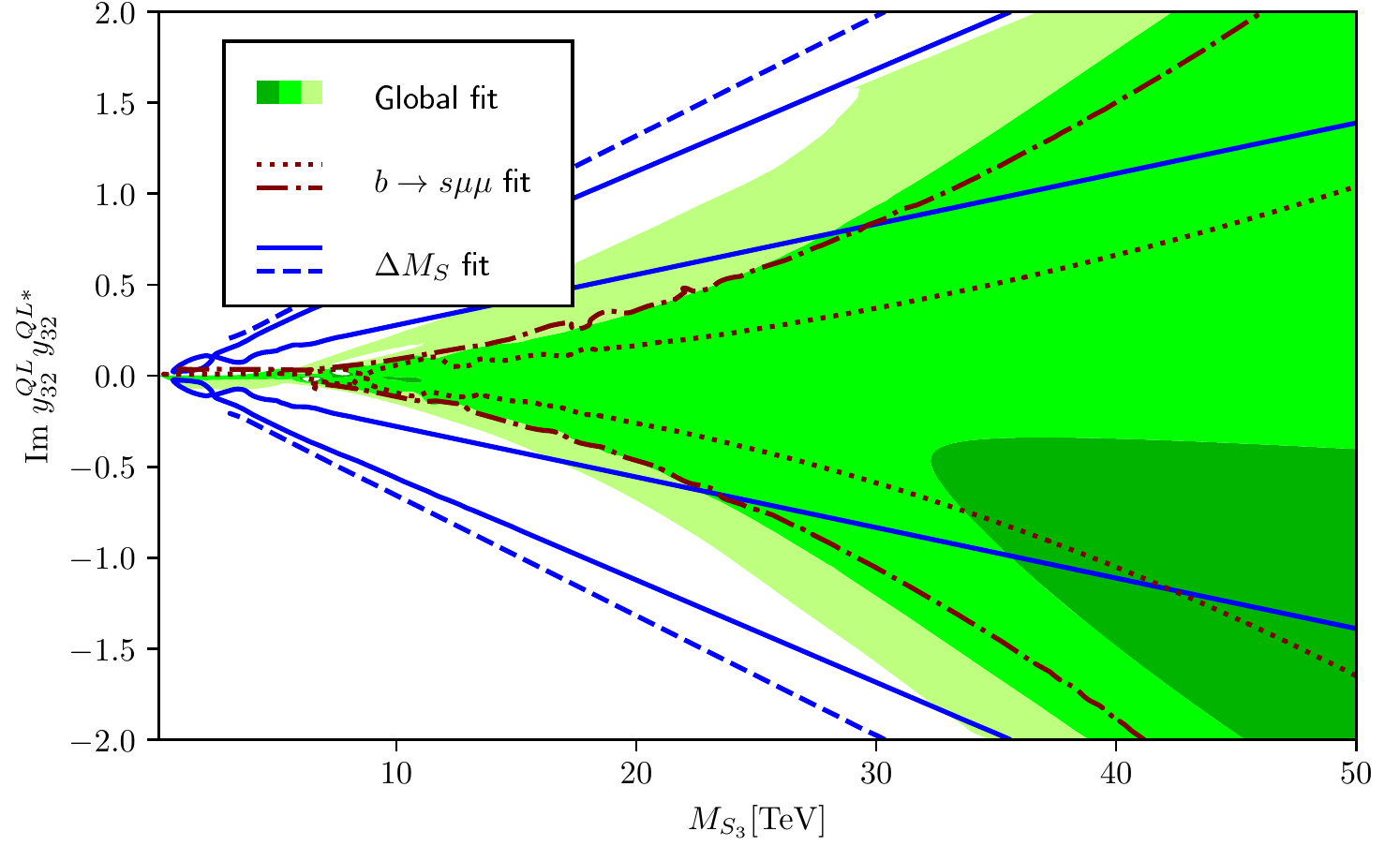}}
\caption{Fit on $S_3$ leptoquark parameter space in the
  $M_{S_3}$-Im $y^{QL}_{32} y^{QL*}_{22}$ plane (see text).}\label{im:WCLQ}
\end{figure} 

The leptoquark model has three independent couplings
  contributing to $\Dms$~\eqref{CbsLQ}. For the global fits we will
  assume that the dominant coupling is the muon coupling $y^{QL}_{32}
  y^{QL*}_{22}$, which is the one contributing to $\Rk$, $\Rks$~\eqref{C9C10LQ}. 
The fits on the $S_3$ leptoquark mass $M_{S_3}$ and the imaginary 
coupling $y^{QL}_{32} y^{QL*}_{22}$  imposed by $\bsmumu$ decays and $B_s$-mixing
are presented in Figure~\ref{im:WCLQ}. The observables used in the
respective fits are the same as in Figure~\ref{im:WCZ}. 
The red lines (dotted, dash-dotted) correspond to the fit using only semi-leptonic
$B$-meson decays, i.e. $\bsmumu$ plus the branching 
ratios $\mathrm{BR}(B_s\to \mu^+ \mu^-)$ and $\mathrm{BR}(B^0 \to \mu^+
\mu^-)$, the best fit region is the one between the curves; 
dotted lines: $\Dchit = 1$, dash-dotted lines: $\Dchit= 4$. Blue lines (solid, dashed) correspond to the fit to $B_s$-mixing observables
$\Delta M_s$ and $\ACPmix$. The best fit region
is the one between the lines;  solid lines $\Dchit= 1$, dashed lines $\Dchit = 4$, there are two
  regions with $\Dchit<1$, but between them $\Dchit$ is
  always smaller than 4. 
  The green regions are the combined global fit: dark region
  $\Dchit\leq 1$, medium $\Dchit\leq 4$ and light $\Dchit\leq 9$. 
In the $\bsmumu$ fit the best fit parameters for
  imaginary couplings is $y_{32}^{QL}y_{22}^{QL*}=-0.2\,i$, $M_{S_3}=40.8\TeV$.
The leptoquark fit to $B_s$-mixing
observables has a double minimum, 
located at $M_{S_3}= 44.9\TeV$, $y_{32}^{QL} y_{22}^{QL*} =
\pm2\;i$, 
with a SM pull of $\sqDchitSM = 1.74$ ($\equiv1.22\,\sigma$) and
$\chitdof=0.51$. These points correspond to a value for the Wilson coefficient
of $\Cll=-1.39\times 10^{-4}$.
The global fit, including all observables, and considering only imaginary $y_{32}^{QL}
  y_{22}^{QL*}$ couplings, is located at $M_{S_3} = 50\TeV$, 
$y_{32}^{QL} y_{22}^{QL*} = -1.67\;i$; with a SM pull of only
$\sqDchitSM= 1.1 (\equiv 0.6\,\sigma)$ and a large $\chitdof=2.16$.
Larger $M_{S_3}$ masses provide similar values for the best
  fit couplings, and observable predictions, and the pulls improve slowly. 
The situation is similar than in the $Z'$ case: by allowing larger
$M_{S_3}$ masses the best fit coupling reaches an asymptotic straight
line, where the
contribution to $\Dms$ is constant~\eqref{CbsLQ}, whereas the
contribution to $|\CntmuNP|$~\eqref{C9C10LQ} decreases as
$M_{S_3}^{-1}$, the best fit coupling behaves as $y_{32}^{QL}
y_{22}^{QL*} \simeq i\,(4.43\times 10^{-2}\times M_{S_3}/\TeV)$. 
Table~\ref{fixLQ} shows the best fit parameters for the leptoquark model considered
in this work, corresponding pulls, predictions to the observables $\Rk$, $\Rks$, $\Delta M_s$ 
and $\ACPmix$ and $\chitdof$, considering real,
imaginary and complex Wilson coefficients.
 Table~\ref{fixLQ} shows that only imaginary couplings do not improve
 the results, they cannot explain the $R_{K^{(*)}}$ anomaly.
However, when complex couplings are considered, we found a better global fit 
of $\Rk, \Rks$ observables,  the best global fit parameters emerge at
$M_{S_3} = 4.1\TeV$ and $y_{32}^{QL} y_{22}^{QL*} = 0.033+0.034\;i$, with 
$\sqDchitSM = 5.90$ ($\equiv 5.27\, \sigma$). 
The best fit point $M_{S_3}$ and
  the coupling real part are similar to the real couplings case. The imaginary part of the coupling is
similar to the real part.
The pull with respect the SM is
marginally better in the case of complex couplings ($\sqDchitSM=5.9$
  versus $5.82$), but it actually worsens in units of $\sigma$, since
  the complex coupling fit has one more free parameter. The
  $\chitdof$ is similar in both scenarios.
The predictions for the $B$-meson physics
observables are similar than in the real couplings case.

\begin{table}[t]
\centering
\begin{tabular}{|c|c|c|c|}\hline
Best fits 	&	Real 	&	Imaginary 	&	Complex 	\\\hline
  $y^{QL}_{32} y^{QL*}_{22}$ 	&	$0.04 $	&	  $ -1.67\;i$ 	&	 $0.033+0.034\;i$ 	\\\hline
 $M_{S_3}$ 	&	5.19  TeV 	&	50  TeV 	&	4.10  TeV 	\\\hline
   Pull ($\sqDchitSM$) 	&	5.82  	&	1.10	&	 5.90	\\\hline
  Pull ($\sigma$) 	&	5.47 $\sigma$ 	&	0.60 $\sigma$	&	5.27 $\sigma$ 	\\\hline
 $\chitdof$  & 		1.38	&	2.16 	&	1.39	\\\hline
 $\Rk$ 	&	$0.64 \pm 0.06$	&	$1.00 \pm 0.01$	& $0.62 \pm 0.14$	\\\hline
 $\Rks^{[0.045,1.1]}$ 	&	$0.835 \pm 0.015$	&	$ 0.93\pm 0.02 $	&	$0.84 \pm 0.04$	\\\hline
 $\Rks^{[1.1, 6]}$ 	&	$ 0.66 \pm 0.06$	&	$1.00 \pm 0.01$	&	$0.66 \pm 0.14$	\\\hline
 $\Delta M_s$	&	$20.07 \pm 1.27$ ps${}^{-1}$ 	&  $18.8 \pm 1.7$ ps${}^{-1}$ 	&	 	  $20.0 \pm 1.2$ ps${}^{-1}$	\\\hline
$\ACPmix$	&	$-0.0374 \pm 0.0006$	&	$ -0.039 \pm 0.002$	&	$-0.032 \pm 0.003$	\\\hline

\end{tabular}
\caption{Best fits, and corresponding pulls, to $\Rk, \Rks, \Delta M_s$ 
and $\ACPmix$; considering real,
imaginary and complex Wilson coefficients on the $S_3$
leptoquark. Shown are also the corresponding pulls,
$\chitdof$ and the predictions for
  semi-leptonic decay observables $\Rk$, $\Rks$; $\Dms$ and $\ACPmix$  with $1\,\sigma$ uncertainties.}
\label{fixLQ}
\end{table}
\begin{figure}[t]
\centering
\resizebox{0.6\textwidth}{!}{\includegraphics{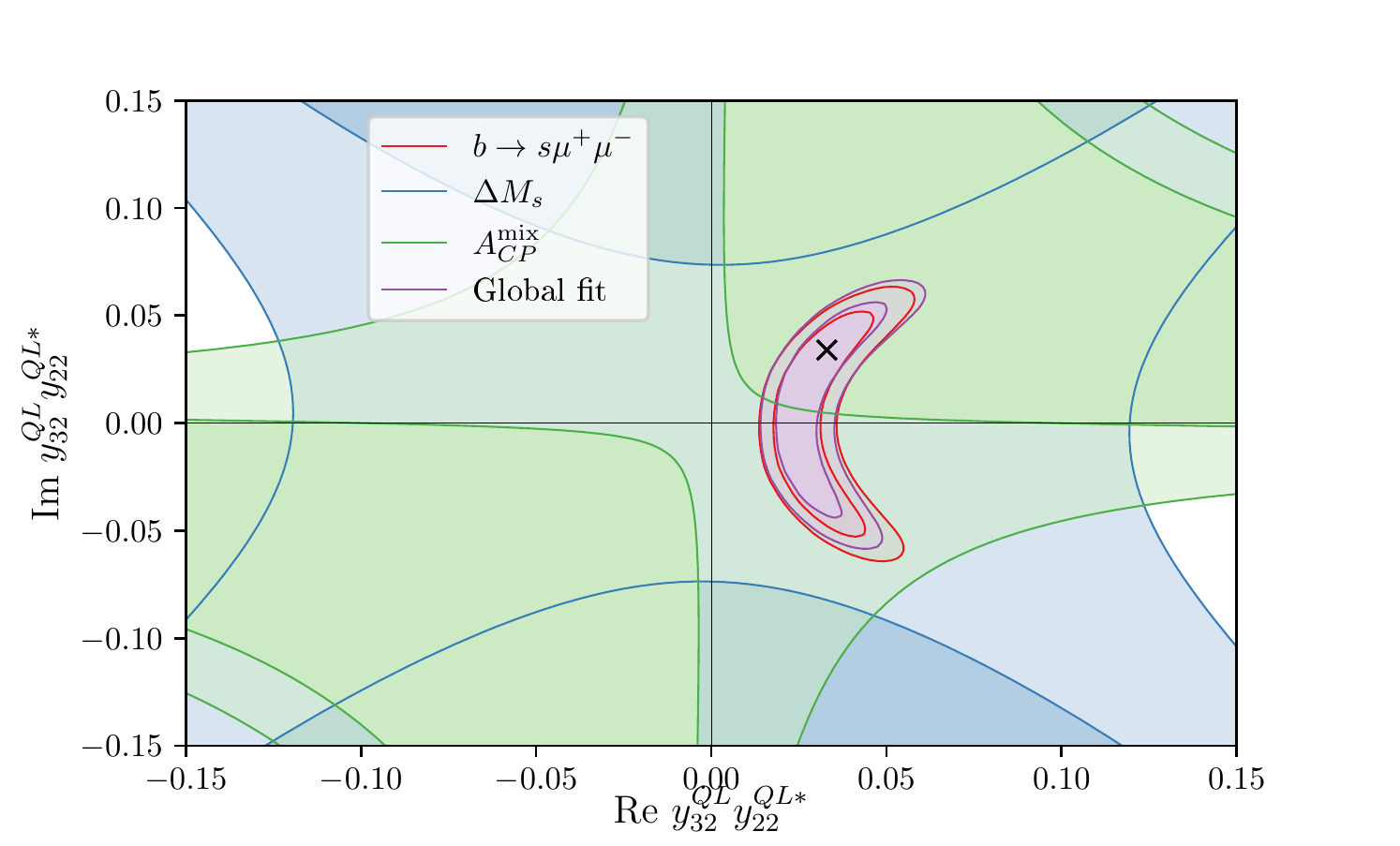}}
\caption{Fit on $S_3$ leptoquark parameter space in the complex $y^{QL}_{32} y^{QL*}_{22}$ plane for the best fit leptoquark
  mass $M_{S_3} =4.1\TeV$ (see text).}
\label{fig:fitcompl_LQ}
\end{figure}

Fig.~\ref{fig:fitcompl_LQ} shows the best fit regions in the
  complex $y_{32}^{QL}  y_{22}^{QL*}$ plane, for the best fit mass
  parameter $M_{S_3}=4.1\TeV$, Table~\ref{fixLQ}. The meaning of each
  region is as in Fig.~\ref{fig:fitcompl_Z}. In this model there is no
  intersection between the 1-$\sigma$ best fit regions of the
  $\bsmumu$ and the $\Dms$ fits. Here we also find the tension between 
  the $\bsmumu$ and $\Dms$ observables, and the different
  evolution of the best fit regions with the leptoquark mass
  $M_{S_3}$. The $\Dms$ fit moves the best fit point away from the
  real axis, and the $\ACPmix$ fit selects of the of the signs for the
  imaginary part, however the global best fit region lies outside the
  1-$\sigma$ region for $\Dms$, and the $\Dms$ prediction does not
  improve with respect the SM.

Ref.\cite{Alok:2017jgr} also provides a fit for the
  leptoquark scenario, our model corresponds to their
  $\vec{\Delta}_{1/3}[S3]$ model. Ref.\cite{Alok:2017jgr} performs a
  fit fixing the leptoquark mass to $M_{S_3}=1\TeV$, and they obtain a
two nearly degenerate minimums with positive and negative imaginary
parts. The reason for that is that they do not include the $\ACPmix$
observable in the fit. Since the $\CnmuNP=-\CtmuNP$ Wilson coefficient
scales like $\sim y_{32}^{QL} y_{22}^{QL*} /M_{S_3}^2$~\eqref{C9C10LQ}
we can compare both results by scaling the best fit coupling with the
mass squared, by taking their central value for the positive imaginary
part, we obtain $y_{32}^{QL} y_{22}^{QL*}=(1.4+1.7\,i)\times 10^{-3}
\times (4.1)^2=0.023+0.029\,i$, which is similar to our third column
in Table~\ref{fixLQ}, and is inside the best fit region of
Fig.~\ref{fig:fitcompl_LQ}. Again, we obtain a larger pull
($\sqDchitSM=5.9$ versus $4.0$).

If one relaxes the condition $y_{33}^{QL} y_{23}^{QL*}\simeq
  y_{31}^{QL} y_{21}^{QL*} \simeq0$ then the leptoquark contributions
  to $\Dms$~\eqref{CbsLQ} and $\CntmuNP$~\eqref{C9C10LQ} are no longer
  correlated, it would be possible to choose: a purely real coupling to
  muons, such that it fulfils the first column of Table~\ref{fixLQ}; a
  vanishing coupling for electrons, such that it does not contribute
  to $\Rk$, $\Rks$; and a complex coupling for taus, such that
  $y_{33}^{QL} y_{23}^{QL*}+y_{32}^{QL} y_{22}^{QL*}$ is purely
  imaginary, and provides a good prediction for
  $\Dms$ like in the second column of Table~\ref{fixLQ}. Of course, this
  would be a quite strange arrangement for leptoquark
  couplings! Another option would be to take an specific model
  construction for the relations among the leptoquark couplings, and
  make a global fit on these parameters. This analysis is beyond the scope of the present work.

\section{Conclusions}
\label{conclusions}

In this work, we have updated the analysis of New Physics 
violating Lepton Flavour Universality, by using the effective Lagrangian
approach and also in the $Z^{'}$ and leptoquark models. 
By considering generic complex Wilson coefficients we found
  that purely imaginary coefficients do not improve significantly
  $B$-meson physics observable predictions, whereas complex coefficients (Table~\ref{tableWC}) do
  improve the predictions, with a slightly improved pull than using only real
  coefficients~\cite{Altmannshofer:2017yso}.
We have analyzed the impact of considering complex Wilson coefficients 
in the analysis of $B$-meson anomalies in two specific models: 
$Z^{'}$ and leptoquarks, and we have presented 
a global fit of $\Rk$ and $\Rks$ observables, together with $\Delta M_s$ and 
$CP$-violation observable $\ACPmix$ when these complex couplings are included
in the analysis. 
We confirm that real Wilson coefficients cannot explain the $B_s$-mixing
anomaly; but also only imaginary Wilson coefficients cannot explain the
$\Rk$, $\Rks$ anomaly. Contrary, complex couplings offer a slightly better global
fit. For complex couplings the predictions for $\Rk$, $\Rks$
  and $\Dms$ are similar than for real couplings (Tables~\ref{tableZ},
\ref{fixLQ}). For $Z'$ models the best fit in both cases is obtained
for $M_{Z'}\simeq 1-1.3\TeV$, a negative real part of the coupling 
${\rm Re}(\lambda_{23}^Q)\simeq -0.002$,
with possibly a similar imaginary coupling part
  ${\rm Im}(\lambda_{23}^Q)\simeq -0.0021$.
For leptoquark models the
situation is similar, with a best fit mass of $M_{S_3}=4-5\TeV$ and a
coupling with a positive real part $y_{32}^{QL} y_{22}^{QL*} \simeq
0.03-0.04$, the presence of a similar imaginary part does not improve
significantly the fit. One can obtain better fits in the
leptoquark models by relaxing the assumption on the leptoquark
couplings, or providing specific models for leptoquark couplings, this
analysis is beyond the scope of the present work. In summary, new physics $Z'$ or leptoquark
models with complex couplings provide a slightly improved global fit
to $B$-meson physics observables as compared with models with real
couplings.

\section*{Note added}
After the completion of this work,
  Ref.~\cite{DiLuzio:2018wch} appeared also analysing the presence of
  complex couplings in the $B$-system. Our results agree with
  Ref.~\cite{DiLuzio:2018wch} wherever comparable.

\section*{Acknowledgments}

J.G. is thankful to F. Mescia for useful discussions.
The work of J.A. and S.P. is partially supported by Spanish
MINECO/FEDER grant FPA2015-65745-P and DGA-FSE grant 2015-E24/2.
S.P. is also supported by CPAN (CSD2007-00042) and FPA2016-81784-REDT.
J.A. is also supported by the 
\textit{Departamento de Innovaci{\'o}n, Investigaci{\'o}n y Universidad} of Aragon
government (DIIU-DGA/European Social Fund).
J.G. has been supported by MCOC (Spain) (FPA2016-76005-C2-2-P),
MDM-2014-0369 of ICCUB (Unidad de Excelencia `Mar{\'\i}a de
Maeztu'), AGAUR (2017SGR754) and CPAN
(CSD2007-00042). J.G. thanks the warm hospitality of the
  Universidad de Zaragoza during the completion of this work.

\end{document}